\documentclass[pra,noshowpacs,english,superscriptaddress,twocolumn]{revtex4-1}
\usepackage[colorlinks,linkcolor=blue,anchorcolor=blue,citecolor=blue,filecolor=blue,menucolor=blue,runcolor=blue,urlcolor=blue,frenchlinks=blue]{hyperref}
\usepackage{amsmath,amssymb,amsthm,amsfonts,mathrsfs}
\usepackage{graphicx}
\usepackage{pdfpages}
\usepackage{appendix}
\usepackage{lipsum}
\usepackage{slashed}
\usepackage{dcolumn}
\usepackage{mathrsfs}
\usepackage{bm}
\usepackage{color}
\usepackage{lineno,hyperref}
\makeatletter

\newcommand{\Rmnum}[1]{\expandafter\@slowromancap\romannumeral #1@}
\AtBeginDocument{\let\LS@rot\@undefined}
\makeatother
\usepackage{tikz,pgf}
\usetikzlibrary{shadings}
\usepackage{pgfplots,pgfplotstable}

\usepackage{tikz,pgf}
\usetikzlibrary{shadings}

\usepackage{pgfplots,pgfplotstable}

\DeclareMathOperator{\sech}{sech}

\newcommand{\rv}{\textcolor{black}}
\begin{document}

\title{Quantum geometric bounds at finite temperature for one-dimensional chiral systems}

\author{Peng He}
\email{hep@swjtu.edu.cn}
\affiliation{ School of Physical Science and Technology, Southwest Jiaotong University, Chengdu 610031, China}

\author{Hai-Tao Ding}
\affiliation{Centre for Quantum Technologies, National University of Singapore, Singapore 117543}
\affiliation{MajuLab, CNRS-UNS-NUS-NTU International Joint Research Unit, Singapore UMI 3654, Singapore}

\author{Yu-Guo Liu}
\email{hugoscholar@outlook.com}
\affiliation{Center on Frontiers of Computing Studies, School of Computer Science, Peking University, Beijing 100871, China}


\date{\today}

\begin{abstract}

The geometry and topology of quantum states are intimately related at zero temperature through exact bounds that constrain geometric quantities from below by topological invariants. At finite temperature, however, the analogous relations remain unclear. Here we establish rigorous geometric lower bounds for one-dimensional chiral-symmetric systems at finite temperature within the Uhlmann's framework for mixed states. We show that the Bures length is bounded by a continuous geometric phase angle. We further derive a temperature-dependent bound that interpolates between the zero-temperature limit and a trivial high-temperature regime. Our results are verified analytically and numerically using the Su-Schrieffer-Heeger model and the spinless Kitaev chain model.  Finally, we discuss potential ways to detect the geometry of the density matrix in quantum circuits, with the quantum imaginary-time evolution method.
\end{abstract}

\maketitle

\section{Introduction}

The geometry and topology of quantum states, encoded in the Berry curvature and the quantum metric, are fundamental concepts underlying a wide range of physical phenomena in condensed matter physics, from the quantization of the Hall conductivity of electronic gases to topological insulators and topological superconductors \cite{DWZhang2018}. Topological phases of matter have traditionally been understood within the framework of pure quantum states at zero temperature. Yet, real quantum systems inevitably reside at finite temperature, where thermal fluctuations render the description in terms of a single wave function inadequate. In recent years, significant efforts have been devoted to extending the concept of topology from pure states to mixed states described by density matrices, based on the Uhlmann phase, a generalization of the Berry phase to mixed quantum states \cite{Carollo2020}. The Uhlmann phase is formulated within the framework of Uhlmann's fiber bundle over the space of density matrices \cite{Uhlmann1986,Uhlmann1991}, and has been used to characterize symmetry-protected topological phases in one-dimensional fermion systems, revealing a temperature-dependent topological phase transition \cite{Viyuela2014}. The notion of finite-temperature topology has been subsequently generalized to two dimensions with the Wilson loop method, which classifies density matrices of two-dimensional fermionic systems and recovers the Chern number in the zero-temperature limit \cite{ZHUang2014,Viyuela2d2014,Viyuela2015,YHe2018}. The finite-temperature topology has since been widely explored in a variety of settings , such as interacting systems \cite{Bardyn2018,Unanyan12020,Molignini2023,ZMHuang2025}, higher-order topological insulators \cite{Chen2026,CLu2025}, and nonequilibrium phase transitions \cite{ZGao2023}.

In parallel, the relation between topology and the quantum metric at zero temperature has been systematically studied both theoretically \cite{CMJian2013,RRoyBand2014,Ozawa2021,Mera2022,Palumbo2018,Zhu2021,AZhang2022,Ding2022,Jankowski2025,JYu2025,PHe2026,Pai2026} and experimentally \cite{asteria2019measuring,XTan2019,MYu2019,gianfrate2020measurement,CRYi2023,TLiu2026}. The quantum metric and the Berry curvature can be unified through the quantum geometric tensor \cite{Provost1980}. For pure states, the volume of the parameter space measured with the quantum metric is found to be bounded from below by the topological invariant, which imposes fundamental constraints on a plethora of geometric phenomena \cite{TLiu2024,JYureview,AGao2025}. For instance, the optical conductivity is directly linked to the quantum geometric tensor via the Souza-Wilkens-Martin sum rule \cite{Niu1985}.  At finite temperature, however, the situation is substantially more complex.  The geometry of mixed states is naturally described by the quantum Bures metric (the real part of the finite-temperature quantum geometric tensor) \cite{Sommers2003,Safronek2017,Carollo2018,ZZhou2024}. Recent work has established finite-temperature generalizations of geometric sum rules, revealing deep connections between thermal response functions and the geometry of the density matrix \cite{GJi2025}. Nevertheless, the precise relationship between topological invariants and geometric quantities at finite temperature remains unclear.

In this work, we address this gap by deriving rigorous quantum geometric bounds for one-dimensional (1D) chiral-symmetric systems at finite temperature. We establish two distinct lower bounds for the Bures length, the natural finite-temperature generalization of the quantum distance \cite{Pai2026,TLiu2026}. We find that the Bures length is bounded from below by the integral of the Uhlmann connection coefficient (which we term the Uhlmann phase angle), rather than directly by the finite-temperature topological invariant, i.e., the Uhlmann phase. Notably, we derive another geometric bound involving the Berry phase and a temperature-dependent damping factor $\tanh (\Delta_{\rm{min}}/T)$, where $\Delta_{\rm{min}}$ is the energy gap and $T$ is the temperature. This bound interpolates between the zero-temperature bound and a trivial bound at high temperatures. Then we numerically verify these bounds in two paradigmatic topological models: the Su-Schrieffer-Heeger (SSH) model \cite{WPSu1979}, and the spinless Kitaev chain model \cite{Kitaev2001}. Specifically, we show that the flat-band solution saturates both the zero-temperature and finite-temperature lower bounds. 

Finally, we outline a verification scheme based on QITE in quantum circuits \cite{Motta2020,Sun2021,Kamakari2022,AnglesCastillo2025}. Applying the general method developed in Ref. \cite{Motta2020}, we obtain an analytical solution of QITE for thermal state preparation on a single qubit. This suggests a potential approach to extract the Bures metric via quantum state tomography, thereby demonstrating that our findings are accessible to quantum simulation platforms.

The rest of this paper is organized as follows. In Sec. \ref{sec2}, we briefly review the Uhlmann's framework of finite-temperature topology and geometry. In Sec. \ref{sec3}, we show the quantum geometric bounds at finite temperature for 1D chiral systems. Then we numerically verify the results by the SSH model and the spinless Kitaev chain model in Sec. \ref{sec4}. In Sec. \ref{sec5}, we propose a verification scheme based on the quantum imaginary-time evolution. Finally, a brief summary is given in Sec. \ref{sec6}.
  
\section{Topology and geometry at finite temperature}\label{sec2}

In this section, we briefly review Uhlmann's framework to treat the mixed-state density matrix on a D-dimensional parameter space. At finite temperature, a quantum system is described by a thermal density matrix $\rho_{\boldsymbol{\lambda}}$ parametrized by $\boldsymbol{\lambda}=(\lambda_1,\lambda_2,\ldots,\lambda_D)$ in D-dimensional parameter space. To define the
topology and geometry for a density matrix of mixed states, we first introduce the amplitude
decomposition or the purification of the density matrix as
\begin{equation}
\rho_{\boldsymbol{\lambda}}=w^\dagger_{\boldsymbol{\lambda}}w_{\boldsymbol{\lambda}},\quad w_{\boldsymbol{\lambda}}=\sqrt{\rho_{\boldsymbol{\lambda}} }U_{\boldsymbol{\lambda}},
\end{equation}
where $w_{\boldsymbol{\lambda}}$ is the amplitude and can be viewed as an analog of the wave function for mixed states, $U_{\boldsymbol{\lambda}} \in U(N)$ represents a gauge degree of freedom generating the same $\rho_{\boldsymbol{\lambda}}$ \cite{Uhlmann1986,Uhlmann1991}. For a smooth closed path $\boldsymbol{\lambda}(t)|_{t=0}^{1}$ in the parameter manifold, Uhlmann introduced a parallel transport condition to lift this $U(n)$ gauge freedom by minimizing the norm of the difference between two neighboring amplitudes $||w_{\boldsymbol{\lambda}+\delta \boldsymbol{\lambda}}-w_{\boldsymbol{\lambda}}||\equiv\operatorname{Tr}[w^\dagger_{\boldsymbol{\lambda}+\delta \boldsymbol{\lambda}} w_{\boldsymbol{\lambda}}]$. For the closed trajectory, initial and final amplitudes are related by a unitary transformation
\begin{equation}
w_{\boldsymbol{\lambda}_1}=w_{\boldsymbol{\lambda}_0}V_{\lambda}.
\end{equation}
Under the Uhlmann parallel transport condition, the $V_{\lambda}$ is a holonomy. Then the Uhlmann phase is given by
\begin{equation}
\Phi_U=\operatorname{arg}(w_{\boldsymbol{\lambda}_1} ,w_{\boldsymbol{\lambda}_0})=\operatorname{arg} \operatorname{Tr}[w^\dagger_{\boldsymbol{\lambda}_0}w_{\boldsymbol{\lambda}_0}V_{\lambda}].
\end{equation}
To visualize the geometric meaning of $\Phi_U$, we can rewrite $\Phi_U$ in form of
\begin{equation}
\Phi_U=\operatorname{arg} \operatorname{Tr}[\rho_{\lambda_0}\mathcal{P}e^{i\oint A}]
\end{equation}
where $\mathcal{P}$ is the path-ordering operator, $\operatorname{arg}[\cdot]=\operatorname{imag}\{\ln [\cdot]\},$ and $A=\sum_\mu A_\mu d\lambda_{\mu}$ is the Uhlmann connection,
\begin{equation}
A_{\mu}=\sum_{ i, j} |\psi_{\boldsymbol{\lambda}}^i\rangle \frac{ \langle\psi_{\boldsymbol{\lambda}}^i | [ (\partial_\mu \sqrt{\rho_{\boldsymbol{\lambda}}} ), \sqrt{\rho_{\boldsymbol{\lambda}}} ] |\psi_{\boldsymbol{\lambda}}^j \rangle}{p_{\boldsymbol{\lambda}}^i+p_{\boldsymbol{\lambda}}^j} \langle\psi_{\boldsymbol{\lambda}}^j | ,\label{eq_uc1}
\end{equation}
in the spectral basis of the density matrix $\rho_{\boldsymbol{\lambda}}=\sum_j p^j_{\boldsymbol{\lambda}}|\psi_{\boldsymbol{\lambda}}^j\rangle\langle \psi_{\boldsymbol{\lambda}}^j|$. \rv{$\Phi_U$ takes quantized values $n\pi$ or $0$ with $n\in \mathbb{Z}$, playing the role of the topological indicator of finite-temperature topological phases.}

On the other hand, to understand the geometry of the density matrix, we consider the Bures distance between $w_{\boldsymbol{\lambda}+\delta \boldsymbol{\lambda}}$ and $w_{\boldsymbol{\lambda}}$ \cite{Carollo2020},
\begin{equation}
ds^2_B \equiv \operatorname{Tr}[ d w^\dagger_{\boldsymbol{\lambda}} dw_{\boldsymbol{\lambda}} ]d\boldsymbol{\lambda}^2,
\end{equation}
where $dw_{\boldsymbol{\lambda}}=G_{\boldsymbol{\lambda}} w_{\boldsymbol{\lambda}}$ with $G_{\boldsymbol{\lambda}}$ being parallel transport generator,
\begin{equation}
\langle \psi_{\boldsymbol{\lambda}}^j| G_{\boldsymbol{\lambda}} |\psi_{\boldsymbol{\lambda}}^k\rangle=\sum_{p^j_{\boldsymbol{\lambda}}, p^k_{\boldsymbol{\lambda}}} \frac{\langle \psi_{\boldsymbol{\lambda}}^j| d \rho|\psi_{\boldsymbol{\lambda}}^k\rangle}{p^j_{\boldsymbol{\lambda}}+p^k_{\boldsymbol{\lambda}}}.
\end{equation}
The Bures metric tensor $g_{\mu\nu}$ is extracted from $ds^2_B=\sum_{\mu\nu} g_{\mu\nu}d\lambda_\mu d\lambda_\nu$,
\begin{equation}
g_{\mu\nu}=\frac{1}{2}\operatorname{Tr}[\{G_{\lambda_\mu},G_{\lambda_\nu}\}\rho_{\boldsymbol{\lambda}}].
\end{equation}
More specifically,
\begin{equation}
\begin{split}
g_{\mu \nu}=&\frac{1}{4} \sum_{p_{\boldsymbol{\lambda}}^k} p^k_{\boldsymbol{\lambda}} \left(\frac{\partial_\mu p^k_{\boldsymbol{\lambda}}}{p^k_{\boldsymbol{\lambda}}}\right)\left(\frac{\partial_\nu p^k_{\boldsymbol{\lambda}}}{p^k_{\boldsymbol{\lambda}}}\right)\\
&+\frac{1}{2} \sum_{\substack{p^j_{\boldsymbol{\lambda}},p^k_{\boldsymbol{\lambda}}}}\left\langle \psi_{\boldsymbol{\lambda}}^j \mid \partial_\mu \psi_{\boldsymbol{\lambda}}^k\right\rangle\left\langle\partial_\nu \psi_{\boldsymbol{\lambda}}^k \mid \psi_{\boldsymbol{\lambda}}^j\right\rangle \frac{\left(p^j_{\boldsymbol{\lambda}}-p^k_{\boldsymbol{\lambda}} \right)^2}{p^j_{\boldsymbol{\lambda}}+p^k_{\boldsymbol{\lambda}}} .\label{eq_bm}
\end{split}
\end{equation}

\section{Geometric bound at finite temperature}\label{sec3}

Now we restrict to the 1D case. At zero temperature, the geometry of a 1D lattice system is described by the Fubini-Study metric of the Bloch states in  momentum space, which is the real part of a quantum geometric tensor,
\begin{equation}
\bar{g}_{kk}=\mathrm{Re}[ \langle\partial_k \psi_k | \partial_k \psi_k \rangle- \langle\partial_k \psi_k | \psi_k \rangle \langle\psi_k | \partial_k \psi_k \rangle].
\end{equation}
The topology of the Bloch states can be indexed by a quantized Berry phase (or Zak phase) $\gamma_B$. Recently, an isoperimetric inequality has been shown \cite{Pai2026},
\begin{equation}
 ( |\gamma_{\mathrm{B}} |-\pi)^2+d_{\mathrm{FS}}^2 \geq \pi^2 ,\label{eq_se0}
\end{equation}
where $d_{\mathrm{FS}}=\int_{\mathrm{BZ}} \sqrt{\bar{g}_{kk}}dk $ is the quantum distance of the first Brillouin zone (BZ). The weak form of the isoperimetric inequality $d_{\mathrm{FS}} \ge |\gamma|$ shows that the quantum distance is bounded by the topological index from below, thus placing fundamental constraints on related geometrical physical quantities. \rv{From the strong inequality, we can see that $d_{\mathrm{FS}}^2 \ge |\gamma_B|(2\pi - |\gamma_B|) \ge |\gamma_B|^2$. The strong inequality thus implies the weak one, but not vice versa. Physically, the strong inequality encodes the detailed geometry of the loop on the Bloch sphere (the relation between the enclosed solid angle and the perimeter), whereas the weak inequality is a global constraint that follows from a pointwise comparison of integrands.}

Let us consider its extension to the mixed states at finite temperature. We focus on 1D chiral-symmetric systems: $S H_k S^{-1}=-H_k$, where $H_k$ is the Bloch Hamiltonian, and $S$ is the symmetry operator. For simplicity, we consider the minimal two-band model. With chiral symmetry, the Bloch Hamiltonian can be built as $H_k=\mathbf{h}_k\cdot \bm{\sigma}=h_x\sigma_x+h_y\sigma_y$, where $\bm{\sigma}=(\sigma_x,\sigma_y,\sigma_z)$ are Pauli matrices. At temperature $T$, the density matrix at a given quasimomentum $k$  can be written as \cite{Viyuela2014,YHe2022}
\begin{equation}
\rho(k)=\frac{1}{2} (I-\mathbf{n}_k\cdot\bm{\sigma} ),\label{eq_dm}
\end{equation}
where $I$ is the $2\times 2$ identity matrix, $\mathbf{n}_k\equiv s_k\hat{\mathbf{h}}_k$ with  $s_k=\tanh (\frac{e_k}{T} )$, $e_k=|\mathbf{h}_k|$, and $\hat{\mathbf{h}}_k=\mathbf{h}_/e_k$.

Plugging Eq. (\ref{eq_dm}) into (\ref{eq_uc1}), we find that the Uhlmann connection takes a simple form (for derivation, see Appendix \ref{appuc}) 
\begin{equation}
A_k^U = i\,\alpha(k)\,\sigma_z,
\end{equation}
where
\begin{equation}
\alpha(k)\equiv -\frac{1}{2}\left(1-\frac{1}{\cosh(e_k/T)}\right)(\hat{h}_x \partial_k \hat{h}_y - \hat{h}_y \partial_k \hat{h}_x).\label{eq_uc2}
\end{equation}
Because the matrices of $A_k^U$ commute with each other for different $k$, the Wilson loop is given by
\begin{equation}
W_k^U = \exp\left(i\sigma_z \oint_{\mathrm{BZ}} \alpha(k)\,dk\right)= \exp(i\Theta\,\sigma_z) ,
\end{equation}
where we define $\Theta \equiv\oint_{\mathrm{BZ}} \alpha(k)dk$ as the Uhlmann phase angle. Then the Uhlmann phase reads as
\begin{equation}
\begin{split}
\Phi_U =& \arg \operatorname{Tr}(\rho_{k_0} W_k^U)\\
=&  \arg \frac{1}{2}\operatorname{Tr}\left[\left(I-\sum_{i=x,y}n_i\sigma_i\right)(\cos \Theta I+i\sin\Theta \sigma_z)\right]\\
=&\arg \cos\Theta .
\end{split}
\end{equation}
\rv{Taking $\Theta \in (-\pi,\pi]$, the Uhlmann phase is $\Phi_U=\pi$ for $|\Theta| \in (\pi/2,\pi]$, and $\Phi^U=0$ for $|\Theta| \in [0,\pi/2)$. In the zero-temperature limit, the thermal factor $1 - \sech(e_k/T)$ in Eq.~(\ref{eq_uc2}) tends to unity, and $\Theta$ reduces to the Berry phase of the occupied band. At finite temperature, $\Theta(T)$ is no longer quantized: the thermal factor $1 - \sech(e_k/T)$ varies continuously with $k$ and $T$, so $\Theta(T)$ interpolates smoothly between its zero-temperature value $\gamma_B$ and zero in high-temperature limit.}

We further calculate the Bures metric as (for derivation, see Appendix \ref{appbm})
\begin{equation}
g_{kk} = \frac{1}{4} \left[ |\partial_k \mathbf{n}_k|^2 + \frac{(\mathbf{n}_k\cdot \partial_k \mathbf{n}_k)^2}{1-|\mathbf{n}_k|^2} \right].\label{eq_gkk}
\end{equation}
To find out the relation between the topology and geometry, we first show a pointwise inequality 
\begin{equation}
g_{kk} \ge |\alpha_k|^2.\label{eq_pi}
\end{equation}
The proof is as follows. From Eq. (\ref{eq_gkk}), we have $g_{kk} \ge \frac{1}{4} \tanh^2(e_k/T)|\partial_k \hat{\mathbf{h}}_k|^2$. From Eq. (\ref{eq_uc2}),
\begin{equation}
|\alpha(k)|^2 = \frac{1}{4}[1-\sech(R/T)]^2 |\partial_k \hat{\mathbf{h}}_k|^2.
\end{equation}
Thus, it suffices to prove $\tanh^2(e_k/T) \ge [1-\sech(e_k/T)]^2$. 
 Indeed,
\begin{equation}
\begin{split}
\tanh x - (1-\sech x) =& \frac{\sinh x}{\cosh x} - 1 + \frac{1}{\cosh x}\\
= &\frac{1-e^{-x}}{\cosh x} \ge 0,
\end{split}
\end{equation}
where we denote $x=e_k/T$ for convenience. Now we arrive at $g_{kk} \ge |\alpha_k|^2$.

With Eq. (\ref{eq_pi}) at hand, we have a geometric bound for the Bures length,
\begin{equation}
L_B= \int_0^{2\pi} \sqrt{g_{kk}}\,dk.
\end{equation}
It is easy to see that $L_B$ is bounded by the Uhlmann phase angle from below (BDI):
\begin{equation}
L_B  \ge \int_0^{2\pi} |\alpha(k)|\,dk \ge |\Theta|.\label{eq_bd2}
\end{equation}
We note that distinct to the zero-temperature case, the Bures length is bounded by the Uhlmann phase angle $\Theta$ rather than the quantized Uhlmann phase $\Phi_U$ itself. Furthermore, we find another bound for the Bures length (BDII),
\begin{equation}
 L_B \ge |\gamma_B| \tanh \left(\frac{\Delta_{\min}}{2T}\right),\label{eq_se}
\end{equation}
where $\Delta_{\min}$ is the minimum gap.

The proof is as follows. The minimum band gap is defined as $\Delta_{\min} = 2\min_k e_k$. Since $e_k \ge \Delta_{\min}/2$, we have $s_k \ge s_{\min}$, with $s_{\min} = \tanh (\frac{\Delta_{\min}}{2T} )$. From Eq. (\ref{eq_g2}), we know
\begin{equation}
g_{kk} \ge \frac{s_{\min}^2}{4} |\partial_k \hat{\mathbf{h}}_k|^2.
\end{equation}
Then,
\begin{equation}
L_B \ge \frac{s_{\min}}{2} \oint_{\mathrm{BZ}} |\partial_k \hat{\mathbf{h}}_k|\,dk = |\gamma_B| \tanh\left(\frac{\Delta_{\min}}{2T}\right).
\end{equation}

This bound bridges the Bures length at finite temperature with the Berry phase at zero temperature. Compared to Eq. (\ref{eq_se0}), the bound at finite temperature contains a damping factor $\tanh (\Delta_{\min}/2T )$. We note that at zero temperature ($T\to 0$): the damping factor tends to 1, and Eq. (\ref{eq_se}) becomes the pure-state weak inequality. In contrast, at high temperature ($T\to\infty$), the right-hand side of Eq. (\ref{eq_se}) vanishes, giving a trivial inequality. \rv{This implies that the finite-temperature geometry retains a memory of the zero-temperature topology, but the memory is exponentially suppressed by temperature, with the minimum gap setting the scale}. 

\rv{So far, we have focused on two-band models. While the two-band case is representative, extending the geometric bounds to multiband systems is desirable for a complete understanding. In general, such an extension is challenging because the Uhlmann connection becomes a genuinely non-Abelian matrix, and the pointwise inequality \(g_{kk} \ge |\alpha(k)|^2\) no longer applies directly. In Appendix \ref{appmultib}, we consider a solvable multiband case in which different band pairs decouple, i.e., no Berry-connection mixing occurs between them. Under this condition, the Uhlmann connection is block diagonal, thus, the bounds can be generalized. This decoupling condition is satisfied in models with a conserved layer or chain index, such as multilayer SSH stacks.}

\section{Applications}\label{sec4}

\begin{figure}[htbp]
	\centering
	\includegraphics[width=0.48\textwidth]{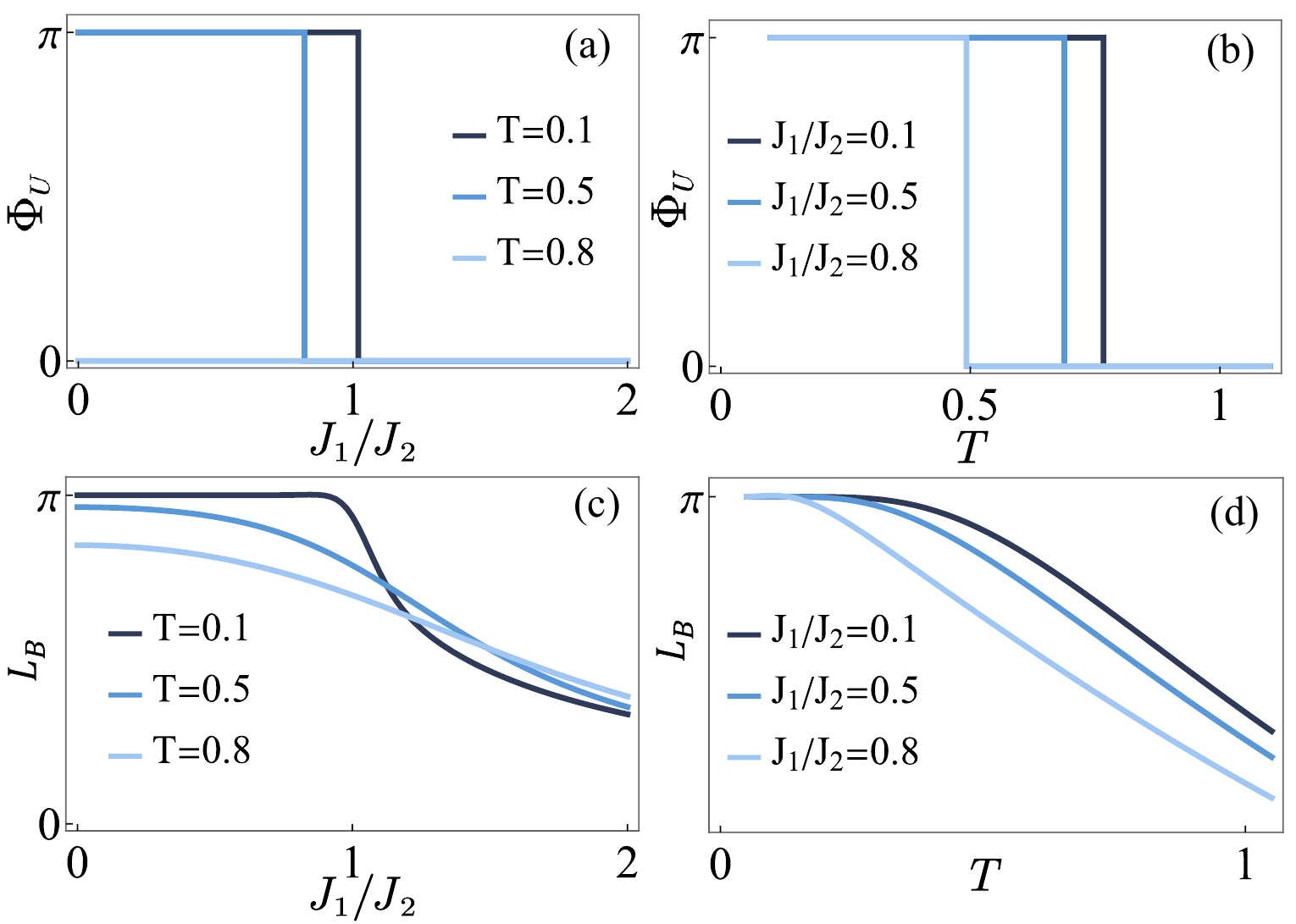}
	\caption{(a) The Uhlmann phase $\Phi_U$ of the SSH model as a function of $J_1/J_2$. (b) The Uhlmann phase $\Phi_U$ of the SSH model as a function of temperature $T$. (c) The Bures length $L_B$ of the SSH model as a function of $J_1/J_2$. (d) The  Bures length $L_B$ of the SSH model as a function of temperature $T$. }
	\label{fig1}
\end{figure}

We use the SSH model and the spinless Kitaev chain to illustrate our finding. The Bloch Hamiltonian of SSH model is given by  $H_{1,k}=(J_1+J_2
\cos k)\sigma_x+J_2\sin k\sigma_y$, where $J_1$ denotes the intra-cell hopping amplitude and $J_2$ the intercell hopping amplitude \cite{WPSu1979}. At zero temperature, the system is topological when $J_1<J_2$.

In Figs. \ref{fig1}(a) and \ref{fig1}(b), we present the phase diagram based on the Uhlmann phase $\Phi_U$ for the SSH model. In the topological regime, the system has a quantized Uhlmann phase $\Phi_U=\pi$. We find that the topological regime reduces as the system undergoes a topological phase transition with raising $T$ (\rv{we note that this is not a thermodynamic phase transition since the free energy and the density matrix evolve smoothly with temperature, and there is no singularity in any thermodynamic quantity}). As indicated by Fig. \ref{fig1}(b), the system sees a higher critical temperature for a larger gap $\Delta_{\min}=2|J_1-J_2|$. At high temperature, the system eventually becomes fully trivial for all parameter regimes. \rv{ It is instructive to estimate the critical temperature \(T_c\) of the finite-temperature topological transition analytically. A low-temperature scaling analysis gives \(T_c \sim \sqrt{\Delta_{\min}}\) (see Appendix \ref{appct}), which is consistent with the numerical results. In contrast, the temperature scale at which BDII becomes trivial is set by the damping factor \(\tanh(\Delta_{\min}/2T)\), which vanishes when \(T \gg \Delta_{\min}/2\), i.e., \(T_{\text{trivial}} \sim \Delta_{\min}/2\). This implies that $T_c>T_{\text{trivial}}$ for vanishing $\Delta_{\min}$, thus, BDII will more quickly become trivial. }

\begin{figure}[htbp]
	\centering
	\includegraphics[width=0.4\textwidth]{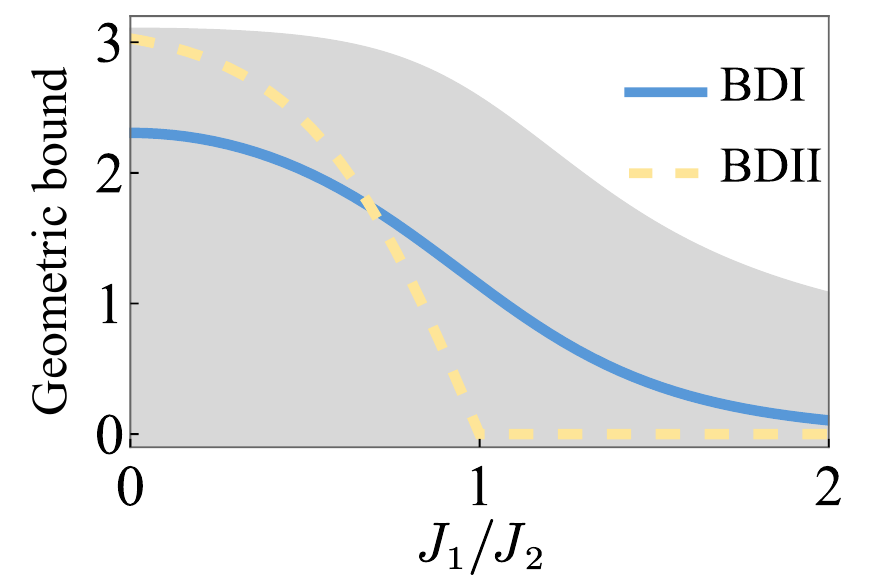}
	\caption{The two geometric bounds for the Bures length of the SSH model at finite temperature $T=0.5$. The gray region shows the values of Bures length.}
	\label{fig2}
\end{figure}

For comparison, we plot the Bures length of the BZ in Figs. \ref{fig1}(c) and \ref{fig1}(d). We observe a larger Bures length for the topological regime. Furthermore, we compare the values of the two geometric bounds for the SSH model with the corresponding Bures length in Fig. \ref{fig2}, taking the $T=0.5$ case as an example. The two inequalities are verified to hold in all parameter regimes. A lower bound is observed for the trivial phase. \rv{The rapid decrease of BDII near the critical point $J_1/J_2=1$ is observed, which is consistent with the scaling analysis.}

\rv{We emphasize a qualitative distinction between the zero-temperature and finite-temperature bounds. At $T=0$,  the geometric bound for the quantum distance is rigid: the geometric quantity is controlled by a discrete topological invariant only taking the values $0$ or $\pi$. At finite temperature, by contrast, the Bures length $L_B$ is bounded from below by the continuous Uhlmann phase angle $\Theta$, rather than by the quantized Uhlmann phase $\Phi^U$ itself. Since $\Theta$ varies continuously with temperature and system parameters, the bound is soft. In particular, this explains why $L_B$ can fall below $\pi$ while $\Phi^U$ remains quantized at $\pi$, which is a phenomenon without zero-temperature counterpart.}

Next, we discuss how to saturate the geometric bound by the flat-band condition, considering the spinless Kitaev chain model, whose BdG Hamiltonian is given by $H_{2,k}=(-\mu-2t\cos k)\sigma_z+2\delta \sin k \sigma_y$ \cite{Kitaev2001}. Importantly, when $\mu=0$ and $\delta=t$, the system features two topological flat (dispersionless) bands: $E=\pm 2t$. At zero temperature, the two flat bands acquire a nonzero Berry phase
\begin{equation}
    \gamma_B=\pm \frac{1}{2} \int_{\mathrm{BZ}}\frac{h_z\partial_k h_y-h_y\partial_k h_z}{h_y^2+h_z^2}dk=\mp  \frac{1}{2} \int_{\mathrm{BZ}} dk=\mp \pi,
\end{equation}
where $-\pi$ corresponds to the lower band and $+\pi$ corresponds to the upper band. And it is easy to see that the quantum metric for the two bands is the same constant number over the full BZ,
\begin{equation}
\overline{g}_{kk}^{\pm}=\frac{1}{2}\operatorname{Tr}[\partial_k P_{\pm}\partial_k P_{\pm}]=\frac{1}{4} (\partial_k \hat{h}_k)^2=\frac{1}{4}.
\end{equation}
The geometric bound obviously saturates in the flat-band limit: $d_{\mathrm{FS}}=|\gamma_B|=\pi$. We can further calculate the Bures metric at the finite temperature, which takes a simple form:
\begin{equation}
g_{kk}=\tanh ^2\left(\frac{2t}{T} \right)|\partial_k \hat{\mathbf{h}}_k|^2.
\end{equation}
Then it is easy to see that the BDII also saturates at the finite temperature. \rv{In fact, the derivation of Eq.~(\ref{eq_bd2}) involves two steps: (i) dropping the radial contribution \(\frac{(s_k')^2}{1-s_k^2}\) and replacing \(s_k\) by its minimum value \(s_{\min}\); (ii) integrating the resulting inequality over the Brillouin zone. Step (i) saturates if and only if \(s_k' = 0\) and \(s_k = s_{\min}\) for all \(k\), which is equivalent to \(e_k = \text{const}\), i.e., a completely flat band. Step (ii) saturates in the topological phase, where \(\frac{1}{2}\int_0^{2\pi}|\partial_k \hat{\mathbf{h}}_k|\,dk = \pi = |\gamma_B|\). Therefore, within the present derivation, BDII is saturated if and only if the band is completely flat and the system is in the topological phase. In the trivial phase (\(\gamma_B = 0\)), BDII reduces to the trivial bound \(L_B \ge 0\), which carries no physical content.}

\begin{figure}[htbp]
	\centering
	\includegraphics[width=0.4\textwidth]{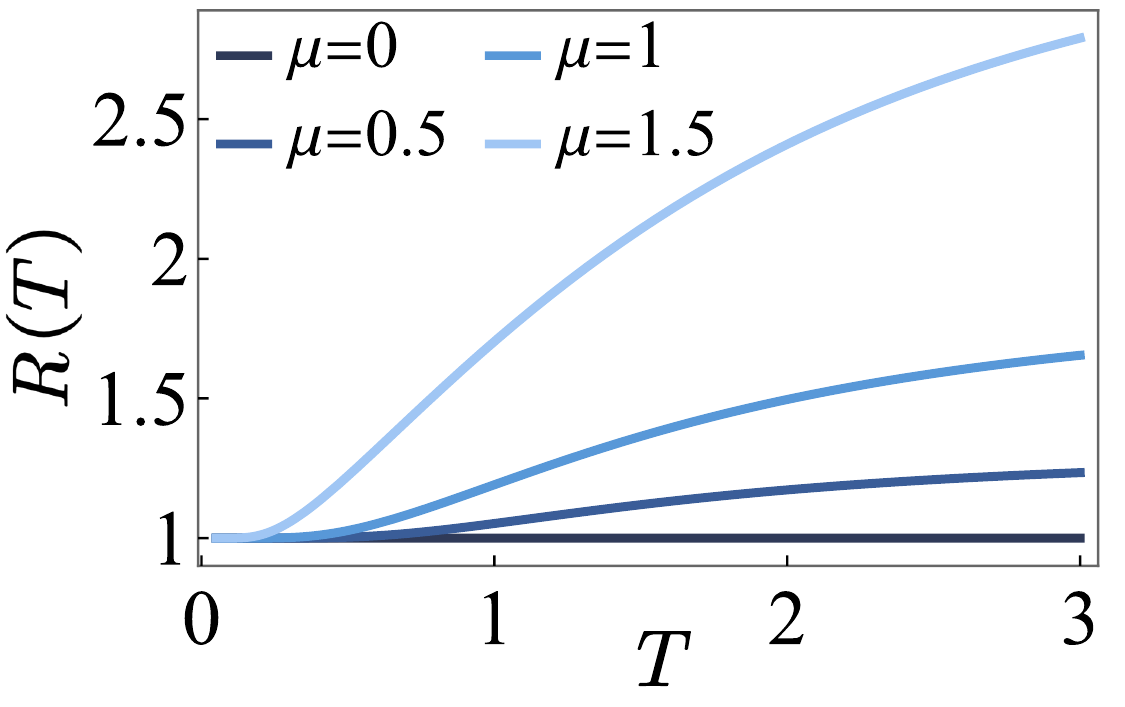}
	\caption{\rv{The tightness ratio $R(T)$ for the Kitaev chain as a function of temperature $T$.}}
	\label{fig3}
\end{figure}

\rv{To quantify the tightness of BDII away from the flat-band limit, we define the tightness ratio
\begin{equation}
R(T) \equiv \frac{\frac{1}{2}\int_0^{2\pi} \tanh(e_k/T)\,|\partial_k \hat{\mathbf{h}}_k|\,dk}{|\gamma_B|\,\tanh(\Delta_{\min}/2T)} \geq 1,
\label{eq:tightness-ratio}
\end{equation}
where the numerator is the tighter bound obtained by retaining the full momentum dependence of \(e_k\) in Eq.~(\ref{eq_g2}), and the denominator is the right-hand side of Eq.~(\ref{eq_bd2}). By construction, \(R(T) = 1\) in the flat-band limit, where BDII is saturated.}

\rv{Figure \ref{fig3} shows \(R(T)\) for the Kitaev chain at several values of \(\mu\) in the topological phase (\(0 < \mu < 2\), with \(\delta = t = 1\)). At \(\mu = 0\), the bands are exactly flat, and \(R(T) = 1\) for all temperatures. As \(\mu\) increases, the bands become dispersive, and \(R(T)\) grows. The growth is more pronounced at higher temperatures and for larger \(\mu\). Near the topological transition (\(\mu =2\)), the dispersion is strongest, and BDII becomes correspondingly weaker. }

\section{Verification via Quantum Imaginary-Time Evolution}\label{sec5}

We now outline a verification scheme based on quantum imaginary-time evolution  in quantum circuits  \cite{Motta2020,Sun2021,Kamakari2022,AnglesCastillo2025}. A detection scheme of the Uhlmann connection has been proposed in Ref. \cite{Viyuela2014}. Here we propose to detect the Bures metric for thermal states. Simulating a finite-temperature quantum system is a generally challenging task. By applying QITE, we prepare the thermal state $\rho(k)$ at each momentum $k$ with inverse temperature $\beta = 1/T$. Through quantum state tomography of the density matrix, we extract the Bures metric $g_{kk}$. The first step is to confirm that the thermal average energy of $\rho(k)$ agrees with that of the exact Gibbs state, thereby validating the QITE procedure. The second step is to verify that the obtained $g_{kk}$ matches the theoretical expression. Once the correct quantum geometry is recovered, the verification of the two inequalities follows naturally as a straightforward consequence.

The key idea of QITE is to use a unitary operator to approximate a short-time, normalized imaginary time evolution:
\begin{equation}
|\psi'\rangle = c^{-1} e^{-\delta\tau H} |\psi\rangle \approx e^{-i\delta\tau M} |\psi\rangle,
\end{equation}
where $c = \sqrt{\langle\psi| e^{-2\delta\tau H} |\psi\rangle}$ is the normalization factor, and $M$ is a Hermitian matrix determined from quantum measurements and classical computation. This unitary operator can be implemented with quantum gates in quantum circuits. For a Bloch Hamiltonian $H_k=(h_x(k), h_y(k), h_z(k))\cdot \bm{\sigma} $, the quantities $M$ and $c$ have analytic forms:
\begin{equation}
M = \frac{\sinh(\theta)}{c \delta \tau} (\hat{\mathbf{h}}_k \times \bm{m})\cdot \bm{\sigma},
\end{equation}
where $\theta=\tau e_k$, $\bm{m}=(m_x, m_y, m_z)$ with $m_{x,y,z}=\langle \psi| \sigma_{x,y,z} | \psi\rangle$, and the normalization factor is given by
\begin{equation}
c=\sqrt{\cosh(2\theta) - \sinh(2\theta) (\hat{\mathbf{h}}_k \cdot \bm{m}) }.
\end{equation}
Details of the derivation are provided in Appendix \ref{AQITE}. 

A standard recipe for preparing a Gibbs state $\rho_{\text{Gibbs}}(\beta)$ at temperature $T = 1/\beta$ is to begin with the infinite-temperature state $\rho(\beta=0) = \frac{1}{2}(|0\rangle\langle0| + |1\rangle\langle1|)$, where $|0\rangle$ and $|1\rangle$ are the eigenstates of $\sigma_z$, and then evolve it under imaginary time for $n = \beta/(2\delta\tau)$ steps. Within the QITE framework, we first prepare two pure states $|\psi_0(\beta/2)\rangle$ and $|\psi_1(\beta/2)\rangle$:
\begin{align}
|\psi_0(\beta/2)\rangle &= \prod_{j=1}^{n} \frac{1}{c_{0j}} e^{-\delta\tau H} |0\rangle = \prod_{j=1}^{n} e^{-i\delta\tau M_{0j}} |0\rangle, \\
|\psi_1(\beta/2)\rangle &= \prod_{j=1}^{n} \frac{1}{c_{1j}} e^{-\delta\tau H} |1\rangle = \prod_{j=1}^{n} e^{-i\delta\tau M_{1j}} |1\rangle,
\end{align}
and subsequently combine them according to the Boltzmann distribution to obtain the density matrix $\rho_{\text{QITE}}(\beta)$, i.e.,
\begin{equation}
\begin{split}
\rho_{\text{QITE}}(\beta) &= P_0 |\psi_0(\beta/2)\rangle\langle\psi_0(\beta/2)| \\
&+ P_1 |\psi_1(\beta/2)\rangle\langle\psi_1(\beta/2)|,
\end{split}
\end{equation}
where $P_{0,1}= \frac{\prod_{j=1}^{n} c_{0,1j}^2}{\prod_{j=1}^{n} c_{0j}^2 + \prod_{j=1}^{n} c_{1j}^2}$.

\begin{figure}[htbp]
	\centering
	\includegraphics[width=0.4\textwidth]{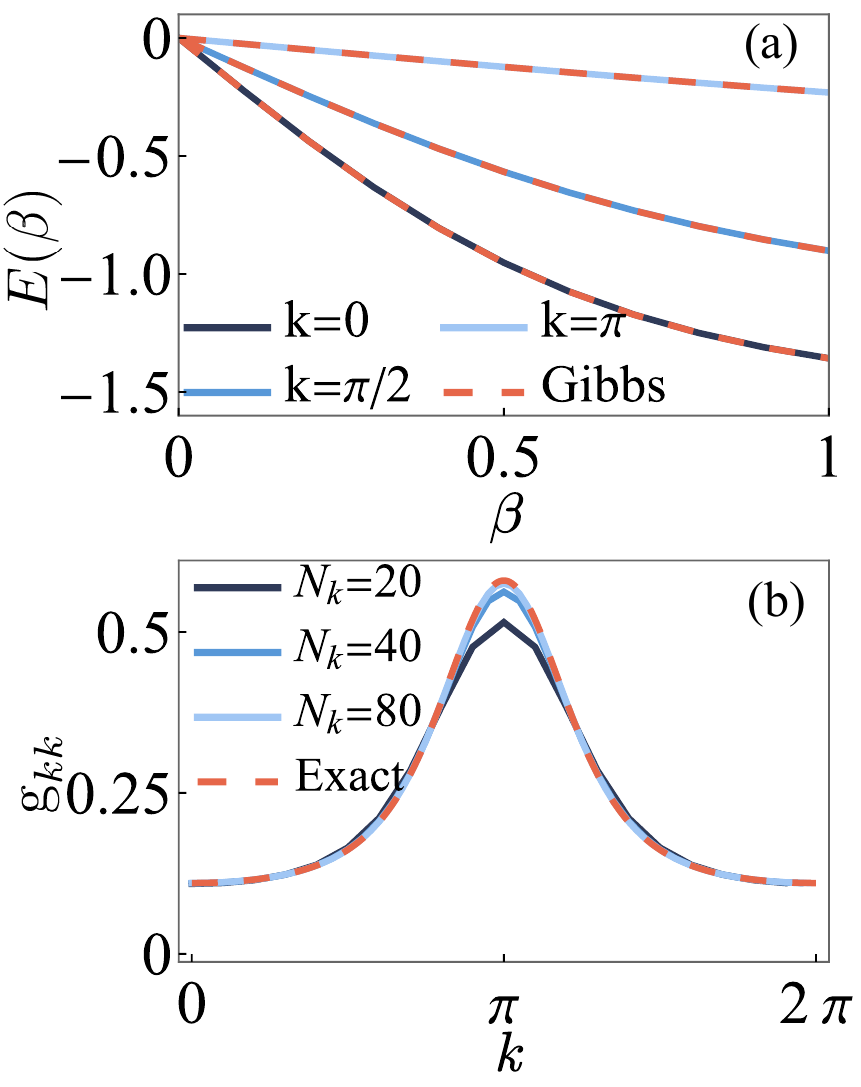}
	\caption{ Numerical demonstration for the Bloch Hamiltonian of the SSH model $H_{1,k}$. (a) Thermal average energy $E(\beta)=\mathrm{Tr}[\rho_{\text{QITE}}(\beta)H_{1,k}]$ as a function of inverse temperature $\beta$ for different momentum sectors $k$ (solid lines), compared with the exact Gibbs state (dashed red line). \rv{(b) Bures metric $g_{kk}$ versus momentum $k$ at $\beta=2$. Solid lines (dark to light) are QITE results for momentum grids with $N_k=20$, $40$, and $80$; the red dashed line is the exact solution. The residual error vanishes as $N_k$ grows. }Both panels are within the topological region ($J_1/J_2=0.5$) with time step $\delta\tau = 0.1$. }
	\label{fig4}
\end{figure}


As a practical demonstration, we perform numerical experiments for the SSH model Hamiltonian $H_{1,k}$. We discretize the Brillouin zone, select sampling momentum points $k$, and employ the QITE method to prepare the target thermal states at each $k$. Since the entire process involves only single-qubit unitary operations and measurements, thermal states of extremely high precision can be obtained. We verify this by extracting the thermal average energy of the target state $E(\beta)=\mathrm{Tr}[\rho_{\text{QITE}}(\beta)H_{1,k}]$ and comparing it with that of the standard Gibbs state $\rho_{\text{Gibbs}}= e^{-\beta H_k}/\mathrm{Tr}(e^{-\beta H_k})$. As shown in Fig. \ref{fig4}(a), this excellent agreement confirms the reliability of the QITE approach for preparing thermal states of single-qubit Hamiltonians. Further analysis in Appendix \ref{errorQITE} shows that our method achieves $O(\delta\tau^2)$ accuracy in thermal state preparation. We scan across different $k$ values and prepare the momentum-resolved thermal states, then measure the average magnetization $\langle \bm{\sigma}\rangle = \operatorname{Tr}\bigl[\rho_{\text{QITE}}(\beta)\,\bm{\sigma}\bigr]=-\bm{n}_k$. The Bures metric $g_{kk}$ can be extracted from $\bm{n}_k$ according to Eq. (\ref{eq_gkk}).  \rv{Here the derivatives entering Eq. (\ref{eq_gkk}) ) are evaluated numerically using central finite differences of   $\bm{n}_k$ on the discretized momentum grid.} We illustrate the extracted Bures metric in Fig.~\ref{fig4}(b), taking the topological region ($J_1=0.5$, $J_2=1$) at a fixed temperature $T=0.5$ ($\beta=2$) as an example. \rv{As the number of momentum sampling points $N_k$ increases, the QITE results steadily approach the exact analytical solutions.} Therefore, this method allows us to identify the geometric structures of thermal states in realistic quantum simulation platforms.

\section{Conclusion}\label{sec6}

In summary, we have established two rigorous geometric bounds for 1D chiral-symmetric systems at finite temperature. Starting from a pointwise inequality between the Bures metric and the Uhlmann connection coefficient, we derived an integrated bound $L_B\ge |\Theta|$, which constrains the Bures length by a continuous geometric phase angle rather than by the quantized Uhlmann phase itself. Building on this result and incorporating the topological winding of the Bloch vector, we further obtained a temperature-damped bound $L_B \ge |\gamma_B| \tanh (\Delta_{\min}/2T)$, which smoothly interpolates between the zero-temperature limit and a trivial high-temperature regime. These bounds have been verified explicitly in the SSH model and the Kitaev chain, and a verification scheme based on quantum imaginary-time evolution has been outlined. \rv{Our findings could hopefully provide a viewpoint for understanding geometry-related responses in thermal mixed states.} Generalization of the present approach to higher-dimensional systems and to other symmetry classes within the Altland-Zirnbauer classification represents a direction for future investigation.

\acknowledgments

P.H. acknowledges support by the Fundamental Research Funds for the Central Universities under Grant No. 2682026CX202.

\begin{appendix}
\section{Calculations of the Uhlmann connection }\label{appuc}
For our two-band system, the density matrix reads as $\rho =\sum_{\pm} p_i P_i$, where $P_{\pm}$ are projectors,
\begin{equation}
P_\pm = |\psi_\pm\rangle\langle \psi_\pm| = \frac{1}{2} (I \pm \hat{\mathbf{h}}\cdot \bm{\sigma}),
\end{equation}
and $p_{\pm}$ are eigenvalues,
\begin{equation}
p_+ = \frac{1+n}{2}, \qquad p_- = \frac{1-n}{2},
\end{equation}
where we denote $n\equiv |\mathbf{n}|=|s_k\hat{\mathbf{h}}_k|=s_k$.
We recall the form of Uhlmann connection,
\begin{equation}
A_k^U = \sum_{i,j} \frac{(\sqrt{p_i}-\sqrt{p_j})^2}{p_i+p_j} |\psi_i\rangle\langle \psi_i| \partial_k |\psi_j\rangle\langle \psi_j|.
\end{equation}
The coefficient for $i\neq j$ is
\begin{equation}
(\sqrt{p_+}-\sqrt{p_-})^2 = 1 - 2\sqrt{p_+p_-} = 1 - \sech(R/T).
\end{equation}
The bracket term simplifies to
\begin{equation}
P_- \partial_k P_+ + P_+ \partial_k P_- = (P_- - P_+) \partial_k P_+ = -(\hat{\mathbf{h}}\cdot \bm{\sigma}) \frac12 (\partial_k \hat{\mathbf{h}} \cdot \bm{\sigma}).
\end{equation}
Direct calculations show that 
\begin{equation}
(\hat{\mathbf{h}}\cdot \bm{\sigma})(\partial_k \hat{\mathbf{h}} \cdot \bm{\sigma}) = i(\hat{h}_x \partial_k \hat{h}_y - \hat{h}_y \partial_k \hat{h}_x)\,\sigma_z,
\end{equation}
where $(\mathbf{a}\cdot \bm{\sigma})(\mathbf{b}\cdot \bm{\sigma}) = (\mathbf{a}\cdot\mathbf{b})I + i(\mathbf{a}\times\mathbf{b})\cdot \bm{\sigma}$ and $\hat{\mathbf{h}}\cdot\partial_k \hat{\mathbf{h}}=0$ are used.
Now we have
\begin{equation}
A_k^U = i\,\alpha(k)\,\sigma_z,
\end{equation}
with
\begin{equation}
\alpha(k) = -\frac{1}{2}\left(1-\frac{1}{\cosh(R/T)}\right)(\hat{h}_x \partial_k \hat{h}_y - \hat{h}_y \partial_k \hat{h}_x).
\end{equation}

\section{Calculations of the Bures metric}\label{appbm}
The Bures metric contains a classical and a quantum contribution. The first term in Eq. (\ref{eq_bm}) is the so-called Fisher-Rao metric, which provides classical probability contribution,
\begin{equation}
\mathcal{T}_1 = \frac{1}{4} \sum_{i=+,-} p_i \left( \frac{\partial_k p_i}{p_i} \right)^2.
\end{equation}
Calculations show that
\begin{equation}
p_+ \left( \frac{\partial_k p_+}{p_+} \right)^2 = \frac{1+n}{2} \left( \frac{\partial_k n}{1+n} \right)^2 = \frac{(\partial_k n)^2}{2(1+n)},
\end{equation}
\begin{equation}
p_- \left( \frac{\partial_k p_-}{p_-} \right)^2 = \frac{1-n}{2} \left( \frac{-\partial_k n}{1-n} \right)^2 = \frac{(\partial_k n)^2}{2(1-n)}.
\end{equation}
Therefore,
\begin{equation}
\mathcal{T}_1 = \frac{1}{4} \left[ \frac{(\partial_k n)^2}{2(1+n)} + \frac{(\partial_k n)^2}{2(1-n)} \right]
= \frac{1}{4} \frac{(\partial_k n)^2}{1-n^2}.\label{eq_t1}
\end{equation}
The second term provides quantum state contribution,
\begin{equation}
\mathcal{T}_2 = \frac{1}{2} \sum_{i\neq j} \langle \psi_i|\partial_k |\psi_j\rangle \langle \psi_j|\partial_k |\psi_i\rangle \frac{(p_i - p_j)^2}{p_i + p_j}.
\end{equation}
Calculations show that
\begin{equation}
\frac{(p_+ - p_-)^2}{p_+ + p_-} = \frac{n^2}{1} = n^2.
\end{equation}
\begin{equation}
|\langle +|\partial_k |-\rangle|^2 = \frac{1}{4} |\partial_k \hat{\mathbf{n}}_k|^2.
\end{equation}
Therefore,
\begin{equation}
\mathcal{T}_2 = \frac{1}{2} \cdot 2 |\langle +|\partial_k |-\rangle|^2 \cdot n^2 = \frac{n^2}{4} |\partial_k \hat{\mathbf{n}}_k|^2.\label{eq_t2}
\end{equation}
Combining Eqs. (\ref{eq_t1}) and (\ref{eq_t2}), we have
\begin{equation}
g_{kk} = \frac{1}{4} \frac{(\partial_k n)^2}{1-n^2} + \frac{n^2}{4} |\partial_k \hat{\mathbf{n}}_k|^2.\label{eq_g1}
\end{equation}
Now express this in terms of $\mathbf{n}_k = n \hat{\mathbf{n}}_k$. Since $\hat{\mathbf{n}}_k\cdot \partial_k \hat{\mathbf{n}}_k = 0$,
\begin{equation}
|\partial_k \mathbf{n}_k|^2 = (\partial_k n)^2 + n^2 |\partial_k \hat{\mathbf{n}}_k|^2,
\end{equation}
and
\begin{equation}
\frac{(\mathbf{n}_k\cdot \partial_k \mathbf{n}_k)^2}{1-|\mathbf{n}_k|^2} = \frac{n^2 (\partial_k n)^2}{1-n^2}.
\end{equation}
Thus Eq. (\ref{eq_g1}) can be rewritten as
\begin{equation}
g_{kk} = \frac{1}{4} \left[ |\partial_k \mathbf{n}_k|^2 + \frac{(\mathbf{n}_k\cdot \partial_k \mathbf{n}_k)^2}{1-|\mathbf{n}_k|^2} \right].
\end{equation}
Substitute $\mathbf{n}_k=s_k\hat{\mathbf{h}}_k$, we also have
\begin{equation}
\begin{split}
g_{kk} =& \frac{1}{4} \left[ \frac{(s_k')^2}{1-s_k^2} + s_k^2 |\partial_k \hat{\mathbf{h}}_k|^2 \right]\\
=&\frac{1}{4}  \left[ \frac{1}{T^2} \sech^2 (\frac{e_k}{T})\cdot (e_k')^2+ \tanh ^2(\frac{e_k}{T} )|\partial_k \hat{\mathbf{h}}_k|^2\right] ,
\end{split}
\label{eq_g2}
\end{equation}
where $s_k'=ds_k/dk$.

\section{\rv{Multiband generalization via band-pair decoupling}}
\label{appmultib}

In this appendix, we generalize the geometric bounds derived in the main text to multiband systems with chiral symmetry. The central obstacle is that for a generic multiband system the Uhlmann connection is a non-Abelian matrix, and the pointwise scalar inequality \(g_{kk} \ge |\alpha(k)|^2\) no longer applies directly. We show that when different band pairs do not mix through the Berry connection \cite{Shem2014}, the multiband problem reduces to \(N\)-independent two-band subsystems, and the geometric bounds generalize accordingly.

We consider a \(2N\)-band system with chiral symmetry \(S H(k) S^{-1} = -H(k)\). In the chiral basis the Hamiltonian takes the block off-diagonal form \(H(k) = \begin{pmatrix} 0 & D(k) \\ D^\dagger(k) & 0 \end{pmatrix}\), where \(D(k)\) is an \(N \times N\) matrix, and the energy spectrum consists of \(N\) pairs due to the symmetry constraint \(\pm \epsilon_a(k)\), \(a = 1, \ldots, N\). We focus on the case where the band pairs decouple, i.e., there exists a \(k\)-independent unitary transformation \(V\) such that \(V D(k) V^\dagger = \operatorname{diag}(d_1(k), \ldots, d_N(k))\) or, equivalently, the projectors \(P_a(k)\) onto the \(a\)th energy pair satisfy \(P_a(k) P_b(k) = \delta_{ab} P_a(k)\) and \(\partial_k P_a = P_a \partial_k P_a + (\partial_k P_a) P_a\). The latter condition means that the matrix elements \(\langle u_i | \partial_k | u_j \rangle\) vanish whenever \(|u_i\rangle\) and \(|u_j\rangle\) belong to different band pairs. In other words, there is no Berry-connection mixing between different band pairs. 
Under this condition, the Hamiltonian block diagonalizes as \(V H(k) V^\dagger = \bigoplus_{a=1}^{N} H_a(k)\) with \(H_a(k) = \mathbf{h}_a(k) \cdot \boldsymbol{\sigma}\), where \(\boldsymbol{\sigma} = (\sigma_x, \sigma_y, \sigma_z)\) acts on the \(a\)th band pair.

For each block, the thermal density matrix at temperature \(T\) is \(\rho_a(k) = \frac{1}{2}[I - \mathbf{n}_a(k) \cdot \boldsymbol{\sigma}]\), where \(\mathbf{n}_a = s_a \hat{\mathbf{h}}_a\), \(s_a = \tanh(e_a/T)\), and \(\hat{\mathbf{h}}_a = \mathbf{h}_a / e_a\). The total density matrix is block diagonal, \(\rho(k) = \bigoplus_a \rho_a(k)\). Because different blocks do not mix, the Uhlmann connection is also block diagonal, \(A_k^U = \bigoplus_{a=1}^{N} i\alpha_a(k) \sigma_z\), where
\begin{equation}
\alpha_a(k) = -\frac{1}{2}\left(1 - \frac{1}{\cosh(e_a / T)}\right)
\left(\hat{h}_{a,x} \partial_k \hat{h}_{a,y} - \hat{h}_{a,y} \partial_k \hat{h}_{a,x}\right).
\end{equation}
Although \(A_k^U\) is a \(2N \times 2N\) matrix, it is block-diagonal with each block proportional to the same fixed generator \(\sigma_z\), so the connection commutes at different momenta: \([A_k^U, A_{k'}^U] = 0\) for all \(k, k'\). The Wilson loop therefore factorizes as \(W^U = \exp(i \sum_{a=1}^{N} \Theta_a \sigma_z^{(a)})\), with \(\Theta_a = \oint_{\mathrm{BZ}} \alpha_a(k) \, dk\).

Since the band pairs are orthogonal, the Bures metric decomposes as \(g_{kk} = \sum_{a=1}^{N} g_{kk}^{(a)}\), where
\begin{equation}
g_{kk}^{(a)} = \frac{1}{4}\left[\frac{(s_a')^2}{1 - s_a^2} + s_a^2 |\partial_k \hat{\mathbf{h}}_a|^2\right].
\end{equation}
For each block, the pointwise inequality derived in the main text holds, \(g_{kk}^{(a)} \ge |\alpha_a(k)|^2\), because \(\tanh x \ge 1 - \sech x\) for all \(x \ge 0\). Summing over all blocks gives \(g_{kk} = \sum_{a=1}^{N} g_{kk}^{(a)} \ge \sum_{a=1}^{N} |\alpha_a(k)|^2\). Define the vector-valued function \(\mathbf{v}(k) = (|\alpha_1(k)|, \ldots, |\alpha_N(k)|)\). Using the vector-valued triangle inequality (also known as Minkowski integral inequality), \(\int_0^{2\pi} \|\mathbf{v}(k)\| \, dk \ge \| \int_0^{2\pi} \mathbf{v}(k) \, dk \| = \sqrt{\sum_{a=1}^{N} (\int_0^{2\pi} |\alpha_a(k)| \, dk)^2}\), together with the scalar triangle inequality \(\int |\alpha_a| \, dk \ge |\int \alpha_a \, dk| = |\Theta_a|\), we obtain the first multiband bound:
\begin{equation}
L_B \ge |\boldsymbol{\Theta}| \equiv \sqrt{\sum_{a=1}^{N} \Theta_a^2}.
\end{equation}
For \(N = 1\) this reduces to the single-band result \(L_B \ge |\Theta|\). For each block, the single-band strong bound gives \(\int_0^{2\pi} |\alpha_a(k)| \, dk \ge |\gamma_{B,a}| \tanh(\Delta_{\min,a}/2T)\), where \(\gamma_{B,a}\) is the Berry phase of the \(a\)-th band pair and \(\Delta_{\min,a} = 2 \min_k \epsilon_a(k)\) is its minimum gap. Substituting into the first bound yields the second multiband bound:
\begin{equation}
L_B \ge \sqrt{\sum_{a=1}^{N} |\gamma_{B,a}|^2 \, \tanh^2\left(\frac{\Delta_{\min,a}}{2T}\right)}.
\end{equation}
When all blocks share the same Berry phase and gap, this simplifies to \(L_B \ge \sqrt{N} \, |\gamma_B| \tanh(\Delta_{\min}/2T)\), exhibiting a \(\sqrt{N}\) enhancement relative to the single-band case. In the multiband decoupled case, the Uhlmann phase becomes \(\Phi^U = \arg \operatorname{Tr}[\rho(k_0) W^U] = \arg(\sum_{a=1}^{N} c_a \cos\Theta_a)\), where \(c_a > 0\) are constants depending on the Boltzmann weights. The Uhlmann phase is therefore a multicomponent object,  which reinforces the central message of the main text: the natural geometric bound involves the continuous phase angles \(\{\Theta_a\}\), not the quantized Uhlmann phase itself.

\section{\rv{Scaling Estimate of the Critical Temperature}}
\label{appct}

\begin{figure}[htbp]
	\centering
	\includegraphics[width=0.4\textwidth]{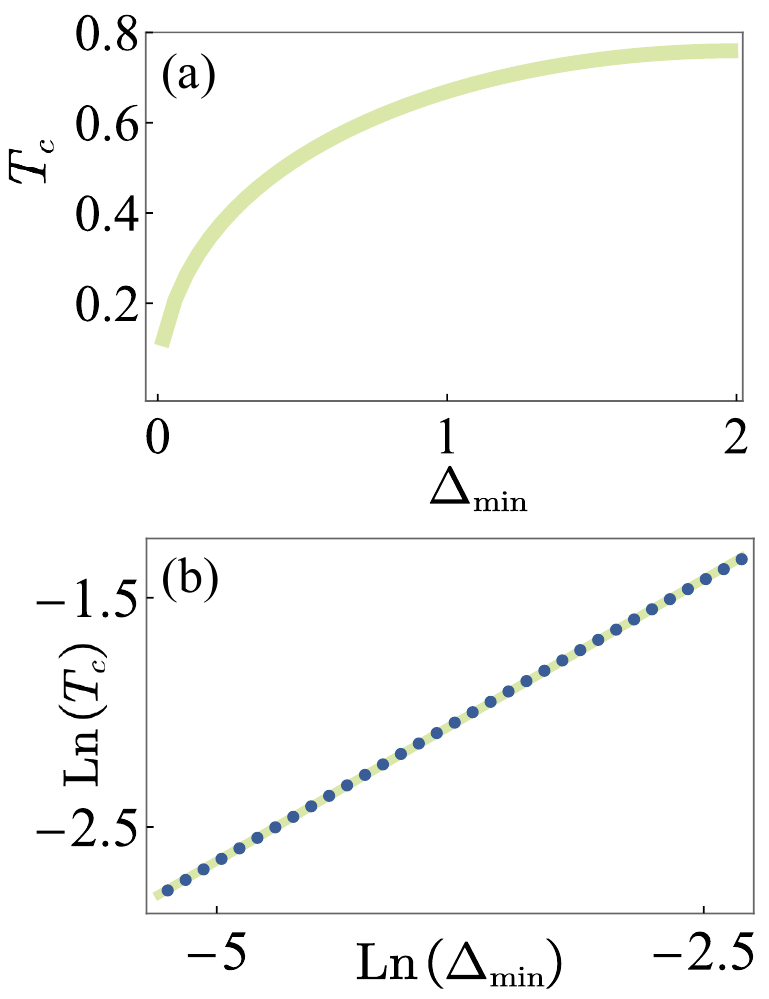}
	\caption{\rv{(a) The critical temperature $T_c$ of the SSH model as a function of the minimum gap $\Delta_{\min}$. (b) The low-temperature window, where $T_c$ scales as the square root of $\Delta_{\min}$. The dots show the numerical data, while the lines show the double-logarithmic fit.}}
	\label{figs}
\end{figure}

In this appendix, we derive the scaling relation between the critical temperature \(T_c\) and the minimum gap \(\Delta_{\min}\) for the SSH model. The critical condition \(\Theta(T_c)=\pi/2\) reads as
\begin{equation}
\int_0^{2\pi}\sech\!\left(\frac{e_k}{T_c}\right)\phi'(k)\,dk=\pi,
\label{eq:Tc-cond}
\end{equation}
where \(e_k=\sqrt{J_1^2+J_2^2+2J_1J_2\cos k}\) and \(\phi'(k)=(J_2^2+J_1J_2\cos k)/e_k^2\). We work in the topological phase \(J_1<J_2\), so that \(\Delta_{\min}=2(J_2-J_1)\) and the minimum of \(e_k\) occurs at \(k=\pi\). Setting \(q=k-\pi\) and expanding \(\cos(\pi+q)\approx-1+q^2/2\), we obtain
\begin{equation}
e_k^2\approx\frac{\Delta_{\min}^2}{4}+J_1J_2\,q^2,~
\phi'(k)\approx\frac{\tfrac{1}{2}J_2\Delta_{\min}+\tfrac{1}{2}J_1J_2\,q^2}{\tfrac{1}{4}\Delta_{\min}^2+J_1J_2\,q^2}.
\label{eq:expansions}
\end{equation}
Here we retain the full \(q\) dependence of \(\phi'(k)\). Introducing the dimensionless variable
\begin{equation}
u\equiv\frac{\sqrt{J_1J_2}}{T_c}\,q,\qquad a\equiv\frac{\Delta_{\min}}{2T_c},
\end{equation}
so that \(q=T_c\,u/\sqrt{J_1J_2}\), \(dq=T_c\,du/\sqrt{J_1J_2}\), and \(e_k/T_c\approx\sqrt{a^2+u^2}\), then the critical condition \eqref{eq:Tc-cond} becomes
\begin{equation}
\pi=\frac{J_2\,a}{\sqrt{J_1J_2}}\,g_1(a)+\frac{\Delta_{\min}}{4a\sqrt{J_1J_2}}\,g_2(a),
\label{eq:main-equation}
\end{equation}
where
\begin{equation}
\begin{split}
 g_1(a)&\equiv\int_{-\infty}^{\infty}\frac{\sech\!\left(\sqrt{a^2+u^2}\right)}{a^2+u^2}\,du,\\
g_2(a)&\equiv\int_{-\infty}^{\infty}\frac{u^2\,\sech\!\left(\sqrt{a^2+u^2}\right)}{a^2+u^2}\,du.   
\end{split}
\end{equation}
Both \(g_1\) and \(g_2\) are positive, even functions of \(a\).

We now analyze Eq.~\eqref{eq:main-equation} in the limit \(\Delta_{\min}\to0\) without assuming \(a\to0\) \textit{a priori}. For small \(a\), the combination \(a\,g_1(a)\) has the expansion
\begin{equation}
a\,g_1(a)=\pi-2a+O(a^2),
\label{eq:ag1}
\end{equation}
where the linear coefficient follows from \(a\,g_1(a)=\int_{-\infty}^{\infty}\sech(a\sqrt{1+x^2})/(1+x^2)\,dx\) and \(\int_{-1/a}^{1/a}dx/(1+x^2)=2\arctan(1/a)\approx\pi-2a\). The function \(g_2(a)\), being even, satisfies
\begin{equation}
g_2(a)=\pi+O(a^2).
\label{eq:g2}
\end{equation}
Substituting Eqs.~\eqref{eq:ag1} and \eqref{eq:g2} into Eq.~\eqref{eq:main-equation}, and using \(\sqrt{J_1J_2}\approx J_2-\Delta_{\min}/4\), we obtain
\begin{equation}
\pi=\pi-2a+\frac{\pi\Delta_{\min}}{4a}+O(a^2,\Delta_{\min}),
\end{equation}
which reduces to the balance condition
\begin{equation}
a^{*2}=\frac{\pi\Delta_{\min}}{8}.
\label{eq:a-star}
\end{equation}
Equation~\eqref{eq:a-star} is self-consistent: it gives \(a^*\propto\sqrt{\Delta_{\min}}\), which indeed vanishes as \(\Delta_{\min}\) approaching zero, justifying the small-\(a\) expansion used to derive it. Consequently,
\begin{equation}
T_c=\frac{\Delta_{\min}}{2a^*}\propto\sqrt{\Delta_{\min}}.
\label{eq:Tc-scaling}
\end{equation}
Therefore, the critical temperature scales as the square root of the minimum gap for small $\Delta_{\min}$.

We numerically confirm the square-root scaling behavior of $T_c$. In Fig \ref{figs}(a), we plot the critical temperature $T_c$ for the SSH model depending on the minimum gap $\Delta_{\min}$.  \(T_c\) increases monotonically with \(\Delta_{\min}\). It rises steeply at small gap regime, and gradually saturates with an increasing gap. In Fig \ref{figs}(b), We perform a double-logarithmic fit to \(T_c\) versus \(\Delta_{\min}\), over the range \(\Delta_{\min}\in[0.005,0.1]\). The extracted slope \( \approx0.4916\) is in good agreement with the analytical prediction \(\frac{1}{2}\).

\section{Analytical solution of QITE for a single-qubit Hamiltonian }\label{AQITE}

We consider the single-qubit Bloch Hamiltonian
\begin{equation}
H_k = \mathbf{h}_k \cdot \boldsymbol{\sigma} = e_k \, \hat{\mathbf{h}}_k \cdot \boldsymbol{\sigma},
\end{equation}
with $e_k = |\mathbf{h}_k|$ and $\hat{\mathbf{h}}_k = \mathbf{h}_k / e_k$. A pure state $|\psi\rangle$ is characterized by its Bloch vector $\mathbf{m} = \langle\psi|\boldsymbol{\sigma}|\psi\rangle$ with $|\mathbf{m}| = 1$. We denote the imaginary-time step by $\delta\tau$ and introduce the parameter $\theta \equiv \delta\tau\, e_k$.

QITE approximates the short-time imaginary-time evolution by a unitary:
\begin{equation}
|\psi'\rangle = c^{-1} e^{-\delta\tau H_k} |\psi\rangle \approx e^{-i\delta\tau M} |\psi\rangle .
\end{equation}
Using $(\hat{\mathbf{h}}_k \cdot \boldsymbol{\sigma})^2 = I$, the imaginary-time propagator expands as
\begin{equation}
\label{eq:prop}
e^{-\delta\tau H_k} = e^{-\theta\,\hat{\mathbf{h}}_k \cdot \boldsymbol{\sigma}}
= \cosh\theta \, I - \sinh\theta \, \hat{\mathbf{h}}_k \cdot \boldsymbol{\sigma},
\end{equation}
from which the normalization factor follows directly:
\begin{equation}
c = \sqrt{\cosh 2\theta - \sinh 2\theta \, (\hat{\mathbf{h}}_k \cdot \mathbf{m})}.
\end{equation}

Following the QITE algorithm of Ref.~\cite{AnglesCastillo2025}, the Hermitian generator $M = \sum_I a_I \sigma_I$ is determined by the linear system
\begin{equation}
(S + S^T) \mathbf{a} = -\mathbf{b},
\label{SAB}
\end{equation}
in the Pauli basis $\sigma_I$ ($I = 0, x, y, z$), where
\begin{equation}
\begin{split}
S_{IJ} &= \langle \sigma_I \sigma_J \rangle, \\
b_I &= \frac{i}{c\,\delta\tau} \big\langle e^{-\delta\tau H_k} \sigma_I - \sigma_I e^{-\delta\tau H_k} \big\rangle .
\end{split}
\end{equation}
For a pure single-qubit state, $S_{ij} = \langle \sigma_i \sigma_j \rangle = \delta_{ij} + i\varepsilon_{ijk} m_k$ for $i,j = x,y,z$, together with $S_{00}=1$, $S_{0i}=m_i$, and $S_{i0}=m_i$. Hence, we obtain
\begin{equation}
S + S^T =
\begin{pmatrix}
2 & 2m_x & 2m_y & 2m_z \\
2m_x & 2 & 0 & 0 \\
2m_y & 0 & 2 & 0 \\
2m_z & 0 & 0 & 2
\end{pmatrix},
\end{equation}
and
\begin{equation}
\mathbf{b} = \begin{pmatrix} b_0 \\ b_x \\ b_y \\ b_z \end{pmatrix}
= -\frac{2\sinh\theta}{c\,\delta\tau}
\begin{pmatrix} 0 \\ \hat h_y m_z - \hat h_z m_y \\ \hat h_z m_x - \hat h_x m_z \\ \hat h_x m_y - \hat h_y m_x \end{pmatrix},
\end{equation}
which is compactly written as 
\begin{equation}
(b_x, b_y, b_z)^T = -\dfrac{2\sinh\theta}{c\,\delta\tau} \, \hat{\mathbf{h}}_k \times \mathbf{m}.
\end{equation}
Solving Eq. (\ref{SAB}) yields $\mathbf{a} = -\mathbf{b}/2$. Consequently,
\begin{equation}
M = \mathbf{a} \cdot \boldsymbol{\sigma} = \frac{\sinh\theta}{c\,\delta\tau} \, (\hat{\mathbf{h}}_k \times \mathbf{m}) \cdot \boldsymbol{\sigma}.
\end{equation}

From a geometric perspective, the generator $M$ rotates the Bloch vector about the instantaneous axis $\hat{\mathbf{h}}_k \times \mathbf{m}$, which is perpendicular to both the Hamiltonian direction and the current magnetization, by an angle
\begin{equation}
\varphi = \delta\tau\,|\mathbf{a}| = \frac{\sinh\theta}{c}\,|\hat{\mathbf{h}}_k \times \mathbf{m}|
= \frac{\sinh\theta}{c}\sqrt{1-(\hat{\mathbf{h}}_k\cdot\mathbf{m})^2}.
\end{equation}
Imaginary time evolution of a single qubit therefore drives the state along the shortest geodesic arc toward the ground state $-\hat{\mathbf{h}}_k$. As $\mathbf{m}$ approaches eigenstates, the rotation angle vanishes ($\varphi \to 0$) and the evolution freezes.

\section{ \rv{Error analysis} } \label{errorQITE}

The error of the QITE algorithm has several distinct sources: the discretization of imaginary time, the Trotter decomposition of the evolution operator, the truncation of the domain on which the unitary is approximated, and the ill-conditioned linear system that solves the generator \textit{M} by a pseudoinverse~\cite{Motta2020} . In this work, however, only the preparation of single-qubit thermal states is required, and the problem admits an exact analytical solution. Consequently, the only algorithmic error that survives is the time-discretization error.

For a single imaginary-time step, comparing the exact evolved state $|\psi'(\delta \tau)\rangle = c^{-1} e^{-\delta\tau H_k} |\psi\rangle$ with the QITE state $|\psi_Q' (\delta \tau)\rangle = e^{-i\delta\tau M} |\psi\rangle $ yields 
\begin{equation}
   \big\lVert\psi'(\delta \tau)-\psi_Q'(\delta \tau)\big\rVert=\frac{\theta^{3}\sin^{3}\alpha}{6}+O(\theta^{4}),
\end{equation}
where $\alpha$ is the angle between the Bloch vector $\bm m$ and $\hat{\bm h}$. This shows that the single-step error of our analytical result, $O(\delta \tau^{3})$, is better than the $O(\delta \tau^{2})$ single-step error of standard QITE ~\cite{Motta2020} .

\begin{figure}[htbp]
	\centering
	\includegraphics[width=0.4\textwidth]{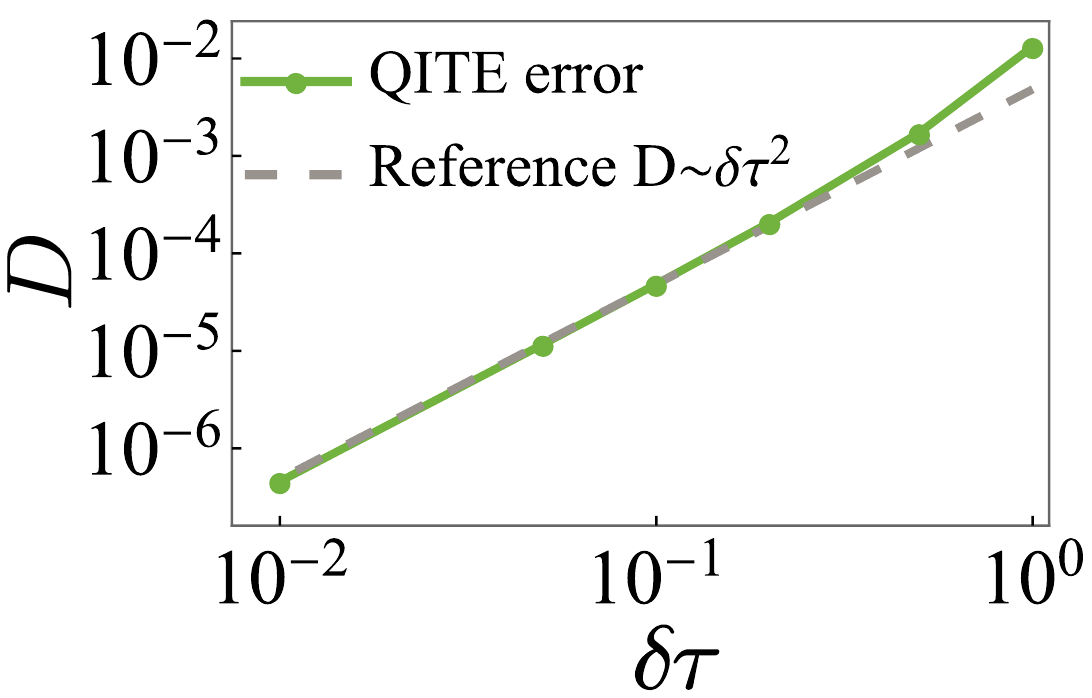}
	\caption{ \rv{Scaling of the thermal-state preparation error with the imaginary-time step $\delta\tau\in[1.0, 0.5, 0.2, 0.1, 0.05, 0.01]$. The trace distance $D$ between the target Gibbs state and the QITE output state follows an $O(\delta\tau^2)$ law (solid line: power-law fit with exponent $\approx2$), verified at $J_1/J_2=0.5$, $\beta=2.0$, $k=0.5\pi$.}}
	\label{fig6}
\end{figure}

For the preparation of the Gibbs thermal state $\rho_G(\beta)=e^{-\beta H}/Z$, we evolve the initial state under imaginary time for $n=\beta/(2\delta\tau)$ steps and obtain $\rho_Q(n)$. During multistep imaginary-time evolution, errors accumulate gradually. To characterize the error scaling of finite-temperature QITE, we keep $\beta$ fixed and sweep $\delta\tau$, using the trace distance $D=\frac{1}{2}\|\rho_G(\beta)-\rho_Q(n)\|_1$ between the target state $\rho_G(\beta)$ and the QITE-prepared state  $\rho_Q(n)$  to quantify the error. The results exhibit a scaling of $O(\delta\tau^2)$ of the thermal-state preparation error, as shown in Fig.~\ref{fig6}.

For a system with $N_B$ bands, $n_q=\lceil\log_2 N_B\rceil$ qubits are required, and the QITE procedure no longer admits an analytical solution. At small $n_q$, full tomography is feasible, requiring $O(4^{n_q})$ measurement settings, with a finite-temperature state-preparation error returning to $O(\delta\tau)$~\cite{Motta2020} and a statistical error decaying as $1/\sqrt{N_{\rm shots}}$.  For large $n_q$, full tomography becomes prohibitively expensive, and instead one employs a truncated QITE unitary together with a Trotter decomposition of the imaginary-time evolution, which lowers the circuit cost at the expense of additional truncation and Trotter errors.

\end{appendix}


\begin{thebibliography}{99}


\bibitem{DWZhang2018}D.-W. Zhang, Y.-Q. Zhu, Y.-X. Zhao, H. Yan, and S.-L. Zhu,
Topological quantum matter with cold atoms, Adv. Phys. \textbf{67}, 253 (2018).

\bibitem{Carollo2020}A. Carollo, D. Valenti, and B. Spagnolo, Geometry of quantum phase transitions, Physics Reports \textbf{838}, 1 (2020).

\bibitem{Uhlmann1986} A. Uhlmann, Parallel transport and “quantum holonomy” along density operators. Reports Math. Phys. 24, 229–240 (1986).

\bibitem{Uhlmann1991}A. Uhlmann, A gauge field governing parallel transport along mixed states, Lett. Math. Phys. \textbf{21}, 229 (1991).

\bibitem{Viyuela2014} O. Viyuela, A. Rivas, and M. A. Martin-Delgado, Uhlmann phase as a topological measure for one-dimensional fermion systems, Phys. Rev. Lett. \textbf{112}(13), 130401 (2014).

\bibitem{ZHUang2014}Z. Huang and D. P. Arovas, Topological indices for open and thermal systems via uhlmann phase,  Phys. Rev. Lett. \textbf{113}, 076407 (2014).

\bibitem{Viyuela2d2014}O. Viyuela, A. Rivas, and M.A. Martin-Delgado, Two-dimensional density-matrix topological fermionic phases: topological Uhlmann numbers, Phys. Rev. Lett. \textbf{113}, 076408 (2014).

\bibitem{Viyuela2015}O. Viyuela, A. Rivas, and M. A. Martin-Delgado, Symmetry-protected topological phases at finite temperature, 2D Mater. \textbf{2}, 034006 (2015).

\bibitem{YHe2018}Y. He, H. Guo, and C. C. Chien, Thermal uhlmann Chern number from the Uhlmann connection for extract
ing topological properties of mixed states, Phys. Rev. B \textbf{97}, 235141 (2018).

\bibitem{Bardyn2018} C. E. Bardyn, L. Wawer, A. Altland, M. Fleischhauer, and S. Diehl, Probing the topology of density matrices, Phys. Rev. X \textbf{8}, 011035 (2018).

\bibitem{Unanyan12020}R. Unanyan, M. Kiefer-Emmanouilidis, and M. Fleischhauer, Finite-temperature topological invariant for interacting systems, Phys. Rev. Lett. \textbf{125}, 215701 (2020).

\bibitem{Molignini2023}P. Molignini and N. R. Cooper, Topological phase transitions at finite temperature, Phys. Rev. Research \textbf{5}, 023004 (2023).

\bibitem{ZMHuang2025}Z.-M. Huang and S. Diehl, Interaction-induced topological phase transition at finite temperature, Phys. Rev. Lett. \textbf{134}, 053002 (2025).

\bibitem{Chen2026}S. Chen and Y. He, The Uhlmann phase of higher-order topological insulators at finite temperature, arXiv: 2606.00479 (2026).

\bibitem{CLu2025}C. Lu, L. Wu, and Q. Ai, Finite-temperature topological invariant for higher-order topological insulators, Phys. Rev. B \textbf{111}, L201101 (2025).

\bibitem{ZGao2023}Z. Gao and Y. He, Quantum quench dynamics of Berry and Uhlmann phases in topological systems, Phys.
Rev. B \textbf{108}, 085126 (2023).

\bibitem{CMJian2013}C.-M. Jian, Z.-C. Gu, and X.-L. Qi, Momentum-space instantons and maximally localized flat
band topological Hamiltonians, Phys. Status Solidi RRL \textbf{7}, 154 (2013).

\bibitem{RRoyBand2014}R. Roy, Band geometry of fractional topological insulators, Phys. Rev. B \textbf{90}, 165139 (2014).

\bibitem{Ozawa2021}T. Ozawa  and B. Mera, Relations between topology and the quantum metric for Chern insulators, Phys. Rev. B \textbf{104}, 045103 (2021).

\bibitem{Mera2022}B. Mera, A. Zhang, and N. Goldman, Relating the topology of Dirac Hamiltonians to quantum
 geometry: When the quantum metric dictates Chern numbers and winding numbers, SciPost Phys. \textbf{12}, 018 (2022).

\bibitem{Palumbo2018} G. Palumbo and N. Goldman, Revealing tensor monopoles through quantum-metric measurements, Phys. Rev. Lett. \textbf{121}, 170401 (2018).

\bibitem{Zhu2021}Y.-Q. Zhu, W. Zheng, S.-L. Zhu, and G. Palumbo, Band topology of pseudo-Hermitian phases through tensor Berry connections and quantum metric, Phys. Rev. B \textbf{104}, 205103 (2021).

\bibitem{AZhang2022} A. Zhang, Revealing Chern number from quantum metric, Chin. Phys. B \textbf{31}, 040201 (2022).

\bibitem{Ding2022} H.-T. Ding, Y.-Q. Zhu, P. He, Y.-G. Liu, J.-T. Wang, D.-W. Zhang, and S.-L. Zhu, Extracting non-Abelian quantum metric tensor and its related Chern numbers, Phys. Rev. A \textbf{105}, 012210 (2022).

\bibitem{Jankowski2025}W. J. Jankowski, A. S. Morris, A. Bouhon1, F. N. Ünal, and R.-J. Slager, Optical manifestations and bounds of topological Euler class, Phys. Rev. B \textbf{111}, L081103 (2025).

\bibitem{JYu2025}J. Yu, J. Herzog-Arbeitman, and B. A. Bernevig, Universal wilson loop bound of quantum geometry: Z2 Bound and physical consequences, Phys. Rev. Lett. \textbf{135}, 086401 (2025).

\bibitem{PHe2026} P. He, J. T. Wang, J. Gong, and H. T. Ding, Floquet quantum geometry in periodically driven topological insulators, arXiv:2602.00569 (2026).

\bibitem{Pai2026}P. Pai and F. Zhang, Isoperimetric inequalities in quantum geometry, Phys. Rev. Lett. \textbf{136}, 116601 (2026)


 \bibitem{asteria2019measuring}L. Asteria, D. T. Tran, T. Ozawa, M. Tarnowski, B. S. Rem, N. Fläschner, K. Sengstock, N. Goldman, and C. Weitenberg,
Measuring quantized circular dichroism in ultracold topological
matter, Nat. Phys. \textbf{15}, 449–454 (2019).

\bibitem{XTan2019} X. Tan, D.-W. Zhang, Z. Yang, J. Chu, Y.-Q. Zhu, D. Li, X. Yang, S. Song, Z. Han, Z. Li, Y. Dong, H.-F.
 Yu, H. Yan, S.-L. Zhu, and Y. Yu, Experimental measurement of the quantum metric tensor and related topological phase transition with a superconducting qubit, Phys. Rev. Lett. \textbf{122}, 210401 (2019).

\bibitem{MYu2019}M. Yu, P. Yang, M. Gong, Q. Cao, Q. Lu, H. Liu, S. Zhang, M. B. Plenio, F. Jelezko, T. Ozawa, N. Goldman, and J. Cai, Experimental measurement of the quantum geometric
 tensor using coupled qubits in diamond, National Science Review \textbf{7}, 254 (2019).

\bibitem{gianfrate2020measurement}A A. Gianfrate, O. Bleu, L. Dominici, V. Ardizzone, M. De
Giorgi, D. Ballarini, G. Lerario, K. W. West, L. N. Pfeiffer,
D. D. Solnyshkov, D. Sanvitto, and G. Malpuech, Measurement of the quantum geometric tensor and of the anomalous Hall drift, Nature (London) \textbf{578}, 381
(2020).

\bibitem{CRYi2023}C.-R. Yi, J. Yu, H. Yuan, R.-H. Jiao, Y.-M. Yang, X. Jiang, J.-Y. Zhang, S. Chen, and J.-W. Pan, Extracting the quantum geometric tensor of an optical Raman lattice by Bloch-state tomography, Phys.
 Rev. Res. \textbf{5}, L032016 (2023).

\bibitem{TLiu2026}T. Liu, Q. Zhang, and C. Qiu, Quantum geometric inequality and its classical wave verification, Phys. Rev. Lett. \textbf{136}, 116602 (2026).

 \bibitem{Provost1980}J. P. Provost and G.Vallee, Riemannian structure on manifolds of quantum states, Commun. Math. Phys. \textbf{76}, 289 (1980).

\bibitem{TLiu2024}T. Liu, X.-B. Qiang, H.-Z. Lu, and X. C. Xie, Quantum geometry in condensed matter, National Science Review \textbf{12}(3), nwae334 (2025).

\bibitem{JYureview} J. Yu, B. A. Bernevig, R. Queiroz,, E. Rossi, P. Törmä, and  B.-J. Yang,  Quantum geometry in quantum materials, npj Quantum Mater. \textbf{10}, 101 (2025). 


\bibitem{AGao2025}A. Gao, N. Nagaosa, N. Ni, and S.-Y. Xu, Quantum geometry phenomena in condensed matter systems, Rev. Mod. Phys. \textbf{98}, 035005 (2026)..

\bibitem{Niu1985}Q. Niu, D. J. Thouless, and Y.-S. Wu, Quantized Hall conductance as a topological invariant, Phys. Rev. B \textbf{31}, 3372 (1985).


\bibitem{Sommers2003} H.-J. rgen Sommers and K. Zyczkowski, Bures volume of the set of mixed quantum states, J. Phys. A: Math. Gen. \textbf{36}(39), 10083 (2003).

\bibitem{Safronek2017}D. Šafránek, Discontinuities of the quantum Fisher information and the Bures metric, Phys. Rev. A \textbf{95}(5), 052320 (2017). 

\bibitem{Carollo2018} A. Carollo, B. Spagnolo, and D. Valenti, Uhlmann curvature in dissipative phase transitions, Scientific reports, \textbf{8}(1), 9852 (2018).

\bibitem{ZZhou2024} Z. Zhou, X. Y. Hou, X. Wang, J. C. Tang, H. Guo, and C. C. Chien,  Sjöqvist quantum geometric tensor of finite-temperature mixed states, Phys. Rev. B, \textbf{110}(3), 035404 (2024).

\bibitem{GJi2025}G. Ji, D. E. Palomino, N. Goldman, T. Ozawa, P. Riseborough, J. Wang, and B. Mera, Density matrix geometry and sum rules, arXiv: 2507.14028 (2025).


\bibitem{WPSu1979}W. P. Su, J. R. Schrieffer, and A. J. Heeger, Phys. Rev. Lett. \textbf{42}, 1698 (1979).

\bibitem{Kitaev2001}A. Y. Kitaev, Unpaired Majorana fermions in quantum wires, Phys.-Usp. \textbf{44}, 131 (2001).

\bibitem{Motta2020}
M. Motta, C. Sun, A. T. K. Tan, M. J. O'Rourke, E. Ye, A. J. Minnich, F. G. S. L. Brandao, and G. K.-L. Chan, Determining eigenstates and thermal states on a quantum computer using quantum imaginary time evolution, Nat. Phys. \textbf{16}, 205--210 (2020).

\bibitem{Sun2021} S.-N. Sun, M. Motta, R. N. Tazhigulov, A. T. K. Tan, G. K.-L. Chan, and A. J. Minnich, Quantum computation of finite-temperature static and dynamical properties of spin systems using quantum imaginary time evolution, PRX Quantum \textbf{2}, 010317 (2021).

\bibitem{Kamakari2022} H. Kamakari, S.-N. Sun, M. Motta, and A. J. Minnich, Digital quantum simulation of open quantum systems using quantum imaginary-time evolution, PRX Quantum \textbf{3}, 010320 (2022).


\bibitem{AnglesCastillo2025}
A. Angl\'{e}s-Castillo, L. Ion, T. Pandit, R. Gomez-Lurbe, R. Martinez, and M. A. Garcia-March,
Understanding quantum imaginary time evolution and its variational form,
in Distributed Computing and Artificial Intelligence, Special Sessions II, 22nd International Conference,
Lecture Notes in Networks and Systems 1630, Springer, Cham (2026), pp. 269--279.


\bibitem{YHe2022}Y. He and C.-C. Chien, Uhlmann holonomy against Lindblad dynamics of topological systems at finite temperatures, Phys. Rev. B \textbf{106}, 024310 (2022).

\bibitem{Shem2014}I. Mondragon-Shem, T. L. Hughes, J. Song, and E. Prodan, Topological criticality in the chiral-symmetric AIII class at strong disorder, Phys. Rev. Lett. \textbf{113}, 046802 (2014).

\end{thebibliography}
\end{document}